\documentstyle[twoside,fleqn,espcrc2,epsf]{article}

\newcommand{\AmS}{{\protect\the\textfont2
  A\kern-.1667em\lower.5ex\hbox{M}\kern-.125emS}}

\title{Lattice fermions with Majorana couplings\thanks{Presented by 
S.~V.~Zenkin}}

\author{S.~Aoki\address{Institute of Physics, University of Tsukuba,
Tsukuba, Ibaraki 305, Japan}, K.~Nagai\address{Center for Computational 
Physics, University of Tsukuba,
Tsukuba, Ibaraki 305, Japan}, and S.~V.~Zenkin\address{Institute for 
Nuclear Research of
the Russian Academy of Sciences, 117312 Moscow, Russia}} 

\begin{document}

\begin{abstract}

We analyse stability of almost massless Dirac mode in gauge models with
boundary (domain wall) fermions, and consider the possibility of decoupling one
of its chiral component by giving it a Majorana mass of the order of the
inverse lattice spacing. We argue that the chiral spectrum in such models is
always uncharged, so they can be implemented for defining the Weyl fermions
only in the real representation of the gauge group, for instance, in SUSY
models.

\end{abstract}

\maketitle

{\bf 1.} Both the Wilson and the domain wall formulations of lattice fermions
involve coupled pair(s) of left-handed ($\psi$) and right-handed ($\chi$)
Weyl fermions. One of the conceivable ways to define within such formulations a
chiral theory, is to decouple, say, $\chi$ by giving it a Majorana mass of the
order of the inverse lattice spacing. It can be done directly, if the fermions
belong to a real representation of the gauge group, or through the Higgs
mechanism, if the representation is complex. In the latter case the model must
have strong coupling paramagnetic (PMS) phase, where fermions acquire masses
$O(1)$ without spontaneous symmetry breaking.

In both formulations the $\psi$ and $\chi$ are coupled through the momentum
dependent Dirac mass terms. Therefore one of the problem on this way is that
the introduction of the Majorana mass for $\chi$ induces a Majorana mass also
for $\psi$, and due to radiative corrections such a mass can get $O(1)$. So,
fine tuning of the Majorana mass of $\psi$ may be necessary. In the Wilson
formulation such a Dirac mass term is the Wilson term $m_D(p) =
\frac{1}{2}\hat{p}^2$ ($\hat{p}_{\mu} = 2 \sin \frac{1}{2} p_{\mu}$), and the
problem of fine tuning of both the Dirac and the Majorana masses does arise.

The boundary fermions \cite{YS} can be viewed as a coupled system of $N_s$ 
pairs of $\psi$ and $\chi$ with the action 

\begin{eqnarray}
&&A_0 = \sum_{s, t = 1}^{N_s}(\overline{\psi}_{s} \delta_{s t} 
D \psi_{t} 
+ \overline{\chi}_{s} \delta_{s t}
\overline{D}\chi_{t} \cr
&&\quad  + \overline{\psi}_{s} W^{-}_{s t} \chi_{t} +
\overline{\chi}_{s} W^{+}_{s t} \psi_{t}),               \label{eq:DMF}
\end{eqnarray}
where 
\begin{eqnarray}
&&D = \nabla_0 + i \sum_{i} \sigma_{i} \nabla_{i}, \quad
\overline{D} = \nabla_0 - i \sum_{i} \sigma_{i} \nabla_{i}, \cr
&&W^{\pm} = \delta_{s \pm 1 \, t} - \delta_{s t}( 1 - M 
- \frac{1}{2}\Delta),               \label{eq:def1}
\end{eqnarray}
and $M \in (0, 2)$ is intrinsic mass parameter of the formulation (for more
detail see \cite{ANZ}). The propagators of such a system have the form
\begin{eqnarray}
&&\langle \psi\overline\psi \rangle_0 = -\overline D G_L \equiv - \overline D 
\frac{1}{\overline{p}^2 + W^{-}W^{+}}, \cr 
&& \langle \chi\overline\chi \rangle_0 = - D G_R \equiv - D 
\frac{1}{\overline{p}^2 + W^{+}W^{-} }, \cr
&&\langle \psi\overline\chi \rangle_0 = W^{-}G_R, \quad 
\langle \chi\overline\psi \rangle_0 = W^{+} G_L, 
\label{eq:prop0}
\end{eqnarray}
with $\overline{p}_{\mu} = \sin p_{\mu}$. The point is that mass matrices 
$W^-W^+$ and $W^+W^-$ have $N_s - 1$
eigenvalues $O(1)$ and exactly one eigenvalue that at $p \sim 0$
have the form $m_{D}^{2} \sim (1-M)^{2 N_s}$,  with the corresponding
eigenvectors 
$\psi \sim \sum_s (1-M)^{s-1} \psi_s$ and $\chi \sim \sum_s (1-M)^{N_s-s} 
\chi_s$.  

Like in the Wilson case, the radiative corrections leads to both a 
multiplicative and an additive renormalization of the mass $M$. However, 
unlike the Wilson fermions, in this case $\delta m_D \neq \delta M$. This 
is illustrated in Fig.~1--2, where numerical results of a mean
field estimate\footnote{In the Feynman gauge the mean field value of the link
variable is $\langle
U_{n, \hat{\mu}} \rangle = \exp[-\frac{1}{2}g^2 C_2(N_c) I]$, where $I=\int_q
(1/\hat{q}^2) \simeq 0.155$ is the tadpole contribution, and $C_2(3) =
4/3$.} of the mass renormalization are shown. 
\begin{figure}[t]
   \vspace*{0cm}
   \epsfysize=5cm
   \epsfxsize=7cm
   \centerline{\epsffile{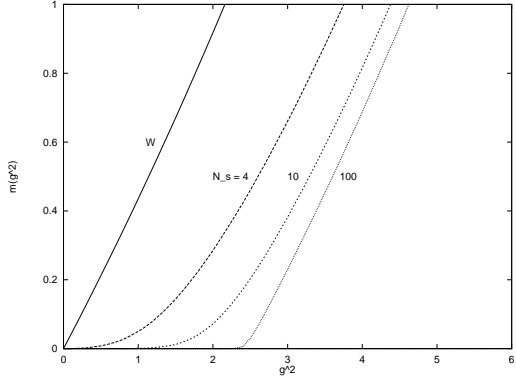}}
   \vspace*{-0cm}
\caption{Mean field estimate of renormalized $m_D$ in QCD as a function of 
$g^2$ at $N_s = 4$, $10$ and $100$, and $M = 0.9$. W stands for the Wilson 
fermions.}
\end{figure}
 
\begin{figure}[t]
   \vspace*{0cm}
   \epsfysize=5cm
   \epsfxsize=7cm
   \centerline{\epsffile{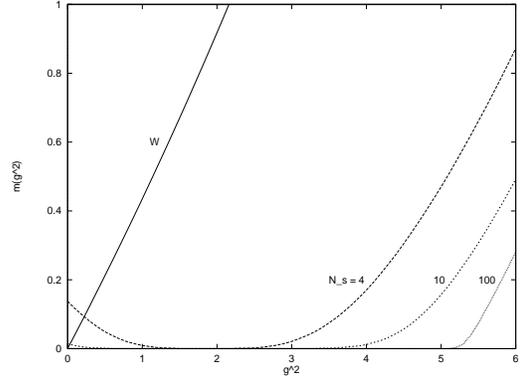}}
   \vspace*{-0cm}
\caption{The same as in Fig.~1 but $M = 1.7$.}
\end{figure}
They demonstrate that for a given
value of the gauge coupling there exists an interval for $M$ with
the midpoint $\overline{M}= 5 - 4 \langle U \rangle(g^2)$ within which
the renormalized $m_D \sim (\overline{M} - M)^{N_s} \ll 1$, 
and the larger $N_s$, the wider this interval. For QCD with $g^2 \leq O(1)$
we have
\begin{equation}
\overline{M} \simeq 1 + 0.4 g^2.
\end{equation}
The situation is not changed with taking into account the nondiagonal parts of
the fermion self energy in perturbation theory (for more detail see
\cite{T})\footnote{See, however, \cite{BS}}. So no fine tuning for the Dirac
mass is needed in such a formulation. This gives rise to a hope that
introducing of the Majorana mass for $\chi$ will not lead to the need for fine
tuning of the Majorana mass of $\psi$.

{\bf 2.} This, in fact, is an underlying motivation of two recent proposals for
the lattice formulations of the Standart Model \cite{CEA} and of $N=1$ SUSY
model \cite{N}. In both proposals it has been suggested to introduce the
Majorana mass only for field $\chi_{N_s}$. Thus the first question arising in
this approach is (i) whether such an introduction of the Majorana mass does
provide a chiral low-lying spectrum and whether the fine tuning is not
needed. Besides, in the case of complex representation of the gauge group
\cite{CEA} the questions of (ii) the existence of the PMS phase and of (iii)
the properties of the system within this phase arise. Here we present the
answers to these questions obtained in \cite{ANZ}.

We consider the system defined by the action 
\begin{equation}
A_m = A_0 + \sum (\chi_{N_s}^{T} H \chi_{N_s} +
\overline{\chi}_{N_s} H^{\dagger} \overline{\chi}_{N_s}^{T}),
\label{eq:DMF1}
\end{equation}
where $H$ either is a constant $H = m \sigma_2$, $m = O(1)$, if the 
representation of 
the gauge group is real, or is a Higgs field $H = y \Phi \sigma_2$
if the representation is complex. So the action (\ref{eq:DMF1}) in both
cases is gauge invariant. 

(i) To answer to the first question consider how the propagators (3)
are modified by the constant Majorana mass term in (5). In the approximation 
neglecting terms $O((1-M)^{N_s})$ the functions $G_{L,R}$ have the form
\begin{eqnarray}
&&G_{L(R)} = A_{L(R)} e^{-\alpha (s+t-2)} \cr
&&\quad \quad \quad + A_{R(L)} e^{\alpha (s+t-2N_s)}
+ B e^{-\alpha |s-t|}, 
\end{eqnarray} 
where $\Re\alpha > 0$, $A_L$, $A_R$ and $B$ are functions of $p$, and only
$A_L$ has a pole at $p^2 = 0$ that corresponds to the massless mode.  The
introduction of the Majorana mass (5) leads to such a modification of the
function $G_{L,R}$, that in $G_L$ only function $A_R$ is modified, so that no
new poles appear in $G_L$, while in $G_R$ only function $A_L$ is modified, and
this modification is such that exactly cancels the pole in $A_L$. Thus only one
(nearly) massless mode of $\psi$ survives in the system (5). Like in the
massless case (3) the main contribution to this mode at low momenta comes from
$\psi_{s=1}$, with exponential dumping of the contributions of higher $s$'s.

It turns out also that at $p \sim 0$ the induced propagator $\langle \psi 
\psi \rangle$ has the form
\begin{equation}
\langle \psi_s \psi_t \rangle \propto p^2 (1-M)^{2N_s+2-s-t},
\end{equation} 
that shows that the effect of the Majorana mass (5) is suppressed 
exponentially in the physical sector\footnote{The expression (7) should be
compared with $\langle \psi \psi \rangle \sim \sigma_2 p^2/(16 m)$ in the 
Wilson case.}. This justifies the hope that no fine tuning of the Majorana 
mass is needed, as well.

(ii) In order that similar situation is realised for complex
representation of the gauge group, the system (5) must be within PMS phase.
That such a phase exists in this system has been demonstrated in 
\cite{ANZ} within a mean field approximation. For instance, in the case of
group U(1) the system is in the PMS phase at $y > 9.7$, and 
$\kappa < \kappa_{cr}(y)$, where $\kappa$ is the standard hopping parameter 
of the Higgs field and $\kappa_{cr}(y) \simeq 1/8 - 11.7/y^2$.

(iii) To get some idea of the properties of the system (5) within the 
PMS phase, we represent the Higgs field $\Phi$ as $\Phi = 
\widetilde{\Phi}^{T} \widetilde{\Phi}$, and use a mean field technique in 
terms of the link expectation value $z^2 = \langle 
\widetilde{\Phi}^{\dagger}_{n} \widetilde{\Phi}_{n \pm \hat{\mu}}\rangle$,
which is known to be nonzero in the PMS phase, though $\langle
\widetilde{\Phi} \rangle = 0$ (for more detail and references see \cite{ANZ}).
We find three possible scenarios. Namely, at tree level the system may 
consist:

either of one neutral (i.e. singlet under the gauge group)
field $\widetilde{\chi} = z \widetilde{\Phi} \chi_{N_s}$ with the
Majorana mass $m = y/z^2$ and charged $\psi$'s and $\chi$'s with 
naive spectrum;

or of pair of massive neutral fields $\widetilde{\chi} = z \widetilde{\Phi}
\chi_{N_s}$ and $\widetilde{\psi} = z \widetilde{\Phi} \psi_{N_s}$ and the
decoupled system (1) with $N_s-1$ pairs of $\psi$ and $\chi$;

or of $N_s$ pair of the neutral fields $\widetilde{\chi}_s = z \widetilde{\Phi}
\chi_{s}$ and $\widetilde{\psi}_s = z \widetilde{\Phi} \psi_{s}$ described 
by the action (5) with modified term $\widetilde{W}^{\pm}_{s t} = 
\delta_{s \pm 1 \, t}/z^2 
-\delta_{s t}[(1-M)/z^2 + \sum_{\mu}(1/z^2 - \cos p_{\mu})]$ 
and the Majorana mass $m = y/z^2$. 

Which of these possibilities is actually realised depends on the parameters of
the system and the answer to this question requires special investigation. 
Only the third scenario leads to the chiral spectrum, provided the mass $M$ is
chosen properly, i.e. within the interval $4-4z^2 < M < 6-4z^2$. However all
the fermions in this case are neutral, and the crucial question to such a 
model is what are the gauge interactions of such neutral fermions.

{\bf 3.} Thus, we have demonstrated that for the real representations of the
gauge group the boundary fermions with the Majorana mass have chiral spectrum
at tree level, and argued that it is stable at least in perturbation theory. 
So,
such models can be implemented for a non-perturbative formulation of the SUSY
models without problem of fine tuning. As concerns the chiral gauge theories,
our conclusion is less optimistic, since we have found that only neutral states
may have chiral spectrum within the PMS phase, and there is a strong evidence
that in the continuum limit such neutral states become noninteracting (see
\cite{ANZ} for more detail and the references).\\

S.V.Z. is grateful to the Organizing Committee of the ``Lattice '97" and to 
the Russian Academy of Sciences for financial support. His work
was partly supported by the Russian Basic Research Fund under the
grant No. 95-02-03868a.

\end{document}